# Magnetization Dynamics in Synthetic Antiferromagnets with Perpendicular Magnetic Anisotropy

Dingbin Huang[1,*], Delin Zhang[2], Yun Kim[1], Jian-Ping Wang[2], and Xiaojia Wang[1,*]

[1]Department of Mechanical Engineering, University of Minnesota, Minneapolis, MN 55455, USA
[2]Department of Electrical and Computer Engineering, University of Minnesota, Minneapolis, MN 55455, USA

**ABSTRACT:**

Understanding the rich physics of magnetization dynamics in perpendicular synthetic antiferromagnets (p-SAFs) is crucial for developing next-generation spintronic devices. In this work, we systematically investigate the magnetization dynamics in p-SAFs combining time-resolved magneto-optical Kerr effect (TR-MOKE) measurements with theoretical modeling. These model analyses, based on a Landau-Lifshitz-Gilbert approach incorporating exchange coupling, provide details about the magnetization dynamic characteristics including the amplitudes, directions, and phases of the precession of p-SAFs under varying magnetic fields. These model-predicted characteristics are in excellent quantitative agreement with TR-MOKE measurements on an asymmetric p-SAF. We further reveal the damping mechanisms of two precession modes co-existing in the p-SAF and successfully identify individual contributions from different sources, including Gilbert damping of each ferromagnetic layer, spin pumping, and inhomogeneous broadening. Such a comprehensive understanding of magnetization dynamics in p-SAFs, obtained by integrating high-fidelity TR-MOKE measurements and theoretical modeling, can guide the design of p-SAF-based architectures for spintronic applications.

---
[*]Authors to whom correspondence should be addressed: huan1746@umn.edu and wang4940@umn.edu

# 1 INTRODUCTION

Synthetic antiferromagnetic (SAF) structures have attracted considerable interest for applications in spin memory and logic devices because of their unique magnetic configurations [1-3]. The SAF structures are composed of two ferromagnetic (FM) layers anti-parallelly coupled through a non-magnetic (NM) spacer, offering great flexibilities for the manipulation of magnetic configurations through external stimuli (*e.g.*, electric-field and spin-orbit torque, SOT). This permits the design of new architectures for spintronic applications, such as magnetic tunnel junction (MTJ), SOT devices, domain wall devices, skyrmion devices, among others [4-7]. The SAF structures possess many advantages for such applications, including fast switching speeds (potentially in the THz regimes), low offset fields, small switching currents (and thus low energy consumption), high thermal stability, excellent resilience to perturbations from external magnetic fields, and large turnability of magnetic properties [3,8-16].

A comprehensive study of the magnetization dynamics of SAF structures can facilitate the understanding of the switching behavior of spintronic devices, and ultimately guide the design of novel device architectures. Different from a single FM free layer, magnetization dynamics of the SAF structures involves two modes of precession, namely high-frequency (HF) and low-frequency (LF) modes, that result from the hybridization of magnetizations precession in the two FM layers. The relative phase and precession amplitude in two FM layers can significantly affect the spin-pumping enhancement of magnetic damping [17], and thus play an important role in determining the magnetization dynamic behaviors in SAFs. Heretofore, the exchange-coupling strength and magnetic damping constant of SAFs have been studied by ferromagnetic resonance (FMR) [18-21] and optical metrology [22-25]. Most FMR-based experimental studies were limited to SAFs with in-plane magnetic anisotropy (IMA). For device applications, perpendicular magnetic anisotropy (PMA) gives better scalability [3,26]. Therefore, the characteristics of magnetization dynamics of perpendicular SAF (p-SAF) structures are of much value to investigate. In addition, prior studies mainly focused on the mutual spin pumping between two FM layers [22,27,28]. A more thorough understanding of the contributions from various sources, including inhomogeneous broadening [29], remains elusive.

In this paper, we report a comprehensive study of the magnetization dynamics of p-SAFs by integrating high-fidelity experiments and theoretical modeling to detail the characteristic parameters. These parameters describe the amplitude, phase, and direction of magnetization precession of both the HF and LF modes for the two exchange-coupled FM layers in a p-SAF. We conduct all-optical time-resolved magneto-optical Kerr effect (TR-MOKE) measurements [30-33] on an asymmetric p-SAF structure with two different FM layers. The field-dependent amplitude and phase of TR-MOKE signals can be well captured by our theoretical model, which in turn provides comprehensive physical insights into the magnetization dynamics of p-SAF structures. Most importantly, we show that inhomogeneous broadening plays a critical role in determining the effective damping of both HF and LF modes, especially at low fields. We demonstrate the quantification of contributions from inhomogeneous broadening and mutual spin pumping (*i.e.*, the exchange of angular momentum between two FM layers via pumped spin currents) [21] to the effective damping, enabling accurate determination of the Gilbert damping for individual FM layers. Results of this work are beneficial for designing p-SAF-based architectures in spintronic applications. Additionally, this work also serves as a successful example demonstrating that TR-MOKE, as an all-optical metrology, is a powerful tool to capture the magnetization dynamics and reveal the rich physics of complex structures that involve multilayer coupling.

## 2    METHODOLOTY

### 2.1    Sample preparation and characterization

One SAF structure was deposited onto thermally oxidized silicon wafers with a 300-nm $SiO_2$ layer by magnetron sputtering at room temperature (RT) in a six-target ultra-high vacuum (UHV) Shamrock sputtering system. The base pressure is below $5 \times 10^{-8}$ Torr. The stacking structure of the SAF is: $[Si/SiO_2]_{sub}/[Ta(5)/Pd(3)]_{seed}/[Co(0.4)/Pd(0.7)/Co(0.4)]_{FM1}/[Ru(0.6)/Ta(0.3)]_{NM}/$ $CoFeB(1)_{FM2}/[MgO(2)/Ta(3)]_{capping}$. The numbers in parentheses denote the layer thicknesses in nanometers. After deposition, the sample was annealed at 250 °C for 20 minutes by a rapid-thermal-annealing process. The two FM layers are CoFeB and Co/Pd/Co layers, separated by a Ru/Ta spacer, forming an asymmetric p-SAF structure (*i.e.*, two FM layers having different magnetic properties). The $M$-$H_{ext}$ loops were characterized by a physical property measurement system (PPMS) with a vibrating-sample magnetometer (VSM) module. The resulting $M$-$H_{ext}$ loops are displayed in Fig. 1(a). Under low out-of-plane fields ($H_{ext} < 500$ Oe), the total magnetic moments in two FM layers of the SAF stack perfectly cancel out each other: $M_1 d_1 = M_2 d_2$ with $M_i$ and $d_i$ being the magnetization and thickness of each FM layer ($i = 1$ for the top CoFeB layer and $i = 2$ for the bottom Co/Pd/Co layer). The spin-flipping field ($H_f \approx 500$ Oe) in the out-of-plane loop indicates the bilinear interlayer-exchange-coupling (IEC) $J_1$ between the two FM layers: $J_1 = -H_f M_{s,1} d_1 \approx -0.062$ erg cm$^{-2}$ [34]. The values of $M_{s,1}$, $M_{s,2}$, $d_1$, and $d_2$ can be found in Table SI of the Supplemental Material (SM) [35].

### 2.2    Theoretical foundation of magnetization dynamics for a p-SAF structure

The magnetic free energy per unit area for a p-SAF structure with uniaxial PMA can be expressed as [36]:

$$F = -J_1(\mathbf{m}_1 \cdot \mathbf{m}_2) - J_2(\mathbf{m}_1 \cdot \mathbf{m}_2)^2$$

$$+ \sum_{i=1}^{2} d_i M_{s,i} \left[ -\frac{1}{2} H_{k,eff,i}(\mathbf{n} \cdot \mathbf{m}_i)^2 - \mathbf{m}_i \cdot \mathbf{H}_{ext} \right] \tag{1}$$

where $J_1$ and $J_2$ are the strength of the bilinear and biquadratic IEC. $\mathbf{m}_i = \mathbf{M}_i/M_{s,i}$ are the normalized magnetization vectors for individual FM layers ($i = 1, 2$). $d_i$, $M_{s,i}$, and $H_{k,eff,i}$ denote, respectively, the thickness, saturation magnetization, and the effective anisotropy field of the $i$-th layer. $\mathbf{n}$ is a unit vector indicating the surface normal direction of the film. For the convenience of derivation and discussion, the direction of $\mathbf{m}_i$ is represented in the spherical coordinates by the polar angle $\theta_i$ and the azimuthal angle $\varphi_i$, as shown in Fig. 1(b).

The equilibrium direction of magnetization in each layer $(\theta_{0,i}, \varphi_{0,i})$ under a given $\mathbf{H}_{ext}$ is obtained by minimizing $F$ in the $(\theta_1, \varphi_1, \theta_2, \varphi_2)$ space. The magnetization precession is governed by the Landau-Lifshitz-Gilbert (LLG) equation considering the mutual spin pumping between two FM layers [27,37-40]:

$$\frac{d\mathbf{M}_i}{dt} = -\gamma_i \mathbf{M}_i \times \mathbf{H}_{eff,i} + \frac{(\alpha_{0,i} + \alpha_{sp,ii})}{M_{s,i}} \mathbf{M}_i \times \frac{d\mathbf{M}_i}{dt} - \frac{\alpha_{sp,ij}}{M_{s,i}} \mathbf{M}_i \times \left( \mathbf{m_j} \times \frac{d\mathbf{m_j}}{dt} \right) \times \mathbf{M}_i \tag{2}$$

On the right-hand side of Eq. (2), the first term describes the precession with the effective field $\mathbf{H}_{eff,i}$ in each layer, given by the partial derivative of the total free energy in the $\mathbf{M}$ space via $\mathbf{H}_{eff,i} = -\nabla_{\mathbf{M}_i} F$. The second term represents the relaxation induced by Gilbert damping ($\alpha$) of the $i$-th layer, which includes the intrinsic ($\alpha_{0,i}$) and spin-pumping-enhanced ($\alpha_{sp,ii}$) damping. For TR-MOKE measurements, $\alpha_{0,i}$ and $\alpha_{sp,ii}$ are indistinguishable. Hence, we define $\alpha_i = \alpha_{0,i} + \alpha_{sp,ii}$ to include both terms. The last term in Eq. (2) considers the influence of pumped spin currents from the layer $j$ on the magnetization dynamics of the layer $i$.

The time evolution of $\mathbf{M}_i$ can be obtained by solving the linearized Eq. (2). Details are provided in Note 1 of the SM [35]. The solutions to Eq. (2) in spherical coordinates are:

$$\begin{bmatrix}\theta_1(t)\\\varphi_1(t)\\\theta_2(t)\\\varphi_2(t)\end{bmatrix}=\begin{bmatrix}\theta_{0,1}\\\varphi_{0,1}\\\theta_{0,2}\\\varphi_{0,2}\end{bmatrix}+\begin{bmatrix}\Delta\theta_1(t)\\\Delta\varphi_1(t)\\\Delta\theta_2(t)\\\Delta\varphi_2(t)\end{bmatrix}=\begin{bmatrix}\theta_{0,1}\\\varphi_{0,1}\\\theta_{0,2}\\\varphi_{0,2}\end{bmatrix}+\begin{bmatrix}C_{\theta,1}^{\mathrm{HF}}\\C_{\varphi,1}^{\mathrm{HF}}\\C_{\theta,2}^{\mathrm{HF}}\\C_{\varphi,2}^{\mathrm{HF}}\end{bmatrix}\exp(i\omega^{\mathrm{HF}}t)+\begin{bmatrix}C_{\theta,1}^{\mathrm{LF}}\\C_{\varphi,1}^{\mathrm{LF}}\\C_{\theta,2}^{\mathrm{LF}}\\C_{\varphi,2}^{\mathrm{LF}}\end{bmatrix}\exp(i\omega^{\mathrm{LF}}t) \qquad (3)$$

with $\Delta\theta_i$ and $\Delta\varphi_i$ representing the deviation angles of magnetization from its equilibrium direction along the polar and azimuthal directions. The last two terms are the linear combination of two eigen-solutions, denoted by superscripts HF (high-frequency mode) and LF (low-frequency mode). $\omega$ is the complex angular frequencies of two modes, with the real and imaginary parts representing the precession angular frequency ($f/2\pi$) and relaxation rate ($1/\tau$), respectively. For each mode, the complex prefactor vector $[C_{\theta,1}, C_{\varphi,1}, C_{\theta,2}, C_{\varphi,2}]^T$ contains detailed information about the magnetization dynamics. As illustrated in Fig. 1(c), the moduli, $|C_{\theta,i}|$ and $|C_{\varphi,i}|$ correspond to the half cone angles of the precession in layer $i$ along the polar and azimuthal directions for a given mode immediately after laser heating, as shown by $\Delta\theta$ and $\Delta\varphi$ in Figs. 1(b-c). The phase difference between $\Delta\theta_i$ and $\Delta\varphi_i$, defined as $\mathrm{Arg}(\Delta\theta_i/\Delta\varphi_i)=\mathrm{Arg}(C_{\theta,i}/C_{\varphi,i})$ with Arg representing the argument of complex numbers, determines the direction of precession. If $\Delta\theta_i$ advances $\Delta\varphi_i$ by 90°, meaning $\mathrm{Arg}(C_{\theta,i}/C_{\varphi,i})=90°$, the precession is counter-clockwise (CCW) in the $\theta$-$\varphi$ space (from a view against $\mathbf{M}_i$). $\mathrm{Arg}(C_{\theta,i}/C_{\varphi,i})=-90°$, on the contrary, suggests clockwise (CW) precession [Fig. 1(d)]. Further, the argument of $C_{\theta,2}/C_{\theta,1}$ provides the relative phase in two FM layers. $\mathrm{Arg}(C_{\theta,2}/C_{\theta,1})=0°$ corresponds to the precession motions in two FM layers that are in-phase (IP) in terms of $\theta$ for a given mode. While the out-of-phase (OOP) precession in terms of $\theta$ is represented by $\mathrm{Arg}(C_{\theta,2}/C_{\theta,1})=180°$ [Fig. 1(e)]. Given the precession direction in each layer and the phase difference between the two FM layers in terms of $\theta$, the phase difference in terms of $\varphi$ can be automatically determined.

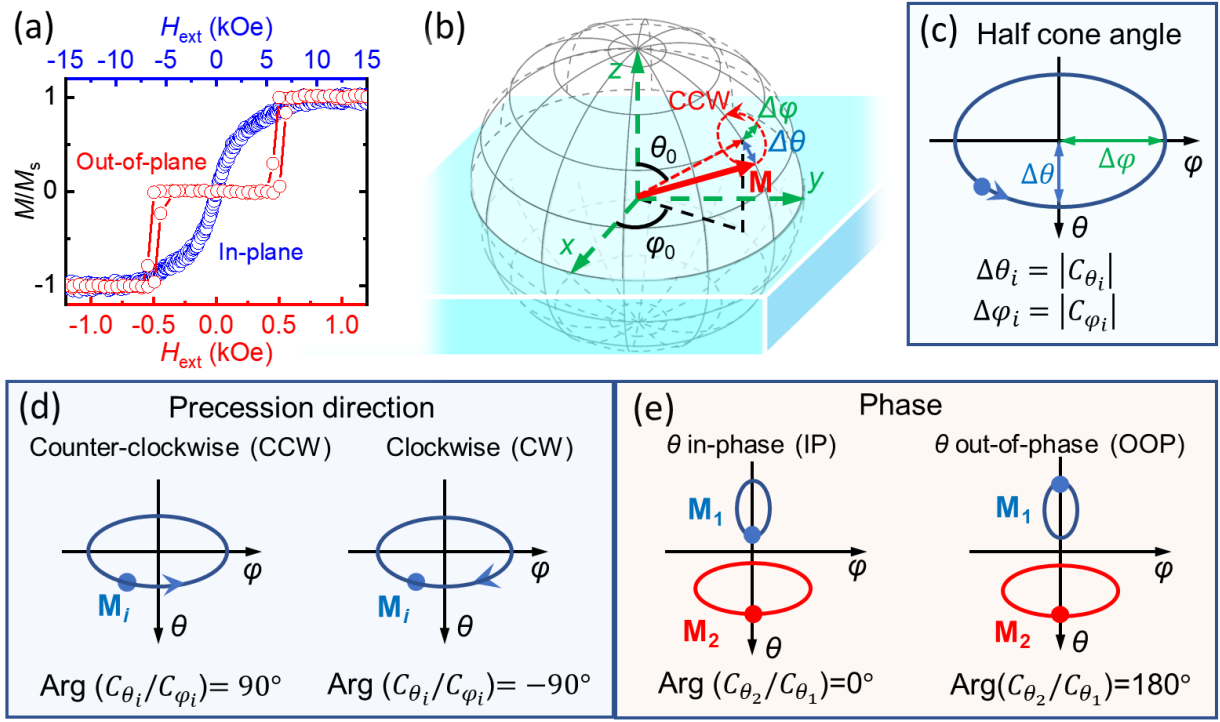

FIG. 1 (a) Magnetic hysteresis ($M$-$H_{\text{ext}}$) loops of the p-SAF stack. The magnetization is normalized to the saturation magnetization ($M/M_s$). (b) Schematic illustration of the half cone angles ($\Delta\theta$ and $\Delta\varphi$) and precession direction of magnetization. The precession direction is defined from a view against the equilibrium direction ($\theta_0$, $\varphi_0$) of $\mathbf{M}$. The representative precession direction in the schematic is counterclockwise (CCW). (c) The relation between precession half cone angles and the prefactors. (d) The relation between precession direction and the prefactors. (e) The relative phase between two FM layers for different prefactor values.

As for the effective damping $\alpha_{\text{eff}} = 1/2\pi f\tau$, in addition to the intrinsic damping ($\alpha_{0,i}$) and the spin-pumping contribution ($\alpha_{\text{sp},ii}$ and $\alpha_{\text{sp},ji}$) considered in Eq. (2), inhomogeneities can also bring substantial damping enhancement [32,33,41,42]. Here, we model the total relaxation rate as follows:

$$\frac{1}{\tau^{\Phi}} = -\text{Im}(\omega^{\Phi}) + \frac{1}{\tau_{\text{inhomo}}^{\Phi}} \tag{4}$$

The superscript $\Phi$ = HF or LF, representing either the high-frequency or low-frequency precession modes. $\omega^{\Phi}$ includes both the intrinsic and spin-pumping contributions. The inhomogeneous broadening is calculated as:

$$\frac{1}{\tau_{\text{inhomo}}^{\Phi}} = \sum_i \frac{1}{\pi} \left| \frac{\partial f^{\Phi}}{\partial H_{\text{k,eff},i}} \right| \Delta H_{\text{k,eff},i} + \sum_i \frac{1}{\pi} \left| \frac{\partial f^{\Phi}}{\partial J_i} \right| \Delta J_i \tag{5}$$

where the first summation represents the contribution from the spatial variation of the effective anisotropy field of individual FM layers ($\Delta H_{\text{k,eff},i}$). The second summation denotes the contribution from the spatial fluctuations of the bilinear and biquadratic IEC ($\Delta J_1$ and $\Delta J_2$). According to Slonczewski's "thickness fluctuations" theory, $\Delta J_1$ generates $J_2$ [43,44]. Therefore, the fact that $J_2$ = 0 for our sample suggests that $\Delta J_1$ is sufficiently small, allowing us to neglect the inhomogeneous broadening from the fluctuations of both the bilinear and biquadratic IEC in the following analyses.

## 2.3 Detection of magnetization dynamics

The magnetization dynamics of the p-SAF sample is detected by TR-MOKE, which is ultrafast-laser-based metrology utilizing a pump-probe configuration. In TR-MOKE, pump laser pulses interact with the sample, initiating magnetization dynamics in magnetic layers via inducing ultrafast thermal demagnetization. The laser-induced heating brings a rapid decrease to the magnetic anisotropy fields and IEC [45,46], which changes $\theta_{0,i}$, $\varphi_{0,i}$ and initiates the precession. The magnetization dynamics due to pump excitation is detected by a probe beam through the magneto-optical Kerr effect. In our setup, the incident probe beam is normal to the sample surface (polar MOKE); therefore, the Kerr rotation angle ($\theta_K$) of the reflected probe beam is proportional to the $z$ component of the magnetization [47]. More details about the experimental setup can be found in Refs. [30,32]. For p-SAF, TR-MOKE signals contain two oscillating frequencies that correspond to the HF and LF modes ($f^{\text{HF}} > f^{\text{LF}}$). The signals are proportional to the change in $\theta_{\text{K}}$ and can be analyzed as follows:

$$\Delta\theta_{\text{K}}(t) = A + Be^{-t/\tau^{\text{T}}} + C^{\text{HF}}\cos(2\pi f^{\text{HF}}t + \beta^{\text{HF}})e^{-t/\tau^{\text{HF}}} + C^{\text{LF}}\cos(2\pi f^{\text{LF}}t + \beta^{\text{LF}})e^{-t/\tau} \quad (6)$$

where the exponential term $Be^{-t/\tau^{\text{T}}}$ is related to the thermal background with $\tau^{\text{T}}$ being the time scale of heat dissipation. The rest two terms on the right-hand side are the precession terms with $C$, $f$, $\beta$, and $\tau$ denoting, respectively, the amplitude, frequency, phase, and relaxation time of the HF and LF modes.

After excluding the thermal background from TR-MOKE signals, the precession is modeled with the initial conditions of step-function decreases in $H_{\text{k,eff},i}$ and $J_i$, following the ultrafast laser excitation [48]. This is a reasonable approximation since the precession period (~15-100 ps for $H_{\text{ext}} > 5$ kOe) is much longer than the time scales of the laser excitation (~1.5 ps) and subsequent relaxations among electrons, magnons, and lattice (~ 1-2 ps) [49], but much shorter than the time scale of heat dissipation-governed recovery (~400 ps). With these initial conditions, the prefactors in Eq. (3) can be determined (see more details in Note 1 of the SM [35]).

For our SAF structure, $\theta_{\text{K}}$ detected by the probe beam contains weighted contributions from both the top and bottom FM layers:

$$\frac{\theta_{\text{K}}(t)}{\theta_{\text{K,s}}} = w\cos\theta_1(t) + (1-w)\cos\theta_2(t) \quad (7)$$

where $\theta_{\text{K,s}}$ represents the Kerr rotation angle when the SAF stack is saturated along the positive out-of-plane ($z$) direction. $w$ is the weighting factor, considering the different contributions to the total MOKE signals from two FM layers. $w$ can be obtained from static MOKE measurements [50], which gives $w = 0.457$ (see more details in Note 2 of the SM [35]).

# 3 RESULTS AND DISCUSSION

## 3.1 Field-dependent precession frequencies and equilibrium magnetization directions

TR-MOKE signals measured at varying $H_{ext}$ are depicted in Fig. 2(a). The external field is tilted 15° away from in-plane [$\theta_H = 75°$, as defined by Fig. 2(c)] to achieve larger amplitdues of TR-MOKE signals [51]. The signals can be fitted to Eq. (6) to extract the LF and HF precession modes. The field-dependent precession frequencies of both modes are summarized in Fig. 2(b). For simplicity, when analyzing precession frequencies, magnetic damping and mutual spin pumping are neglected due to its insignificant impacts on precession frequencies. By comparing the experimental data and the prediction of $\omega^{HF}/2\pi$ and $\omega^{LF}/2\pi$ based on Eq. (3), the effective anisotropy fields and the IEC strength are fitted as $H_{k,eff,1} = 1.23 \pm 0.28$ kOe, $H_{k,eff,2} = 6.18 \pm 0.13$ kOe, $J_1 = -0.050 \pm 0.020$ erg cm$^{-2}$, and $J_2 = 0$. All parameters and their determination methods are summarized in Table SI of the SM [35]. The fitted $J_1$ is close to that obtained from the $M$-$H_{ext}$ loops (~$-0.062$ erg cm$^{-2}$). The inset of Fig 2(b) shows the zoomed-in view of field-dependent precession frequencies around $H_{ext} = 8$ kOe, where an anti-crossing feature is observed: a narrow gap (~2 GHz) opens in the frequency dispersion curves of the HF and LF modes owing to the weak IEC between two FM layers. Without any IEC, the precession frequencies of two FM layers would cross at $H_{ext} = 8$ kOe, as indicated by the green dashed line and blue dashed line in the figure. We refer to these two sets of crossing frequencies as the single-layer natural frequencies of two FM layers (FM$_1$ and FM$_2$) in the following discussions.

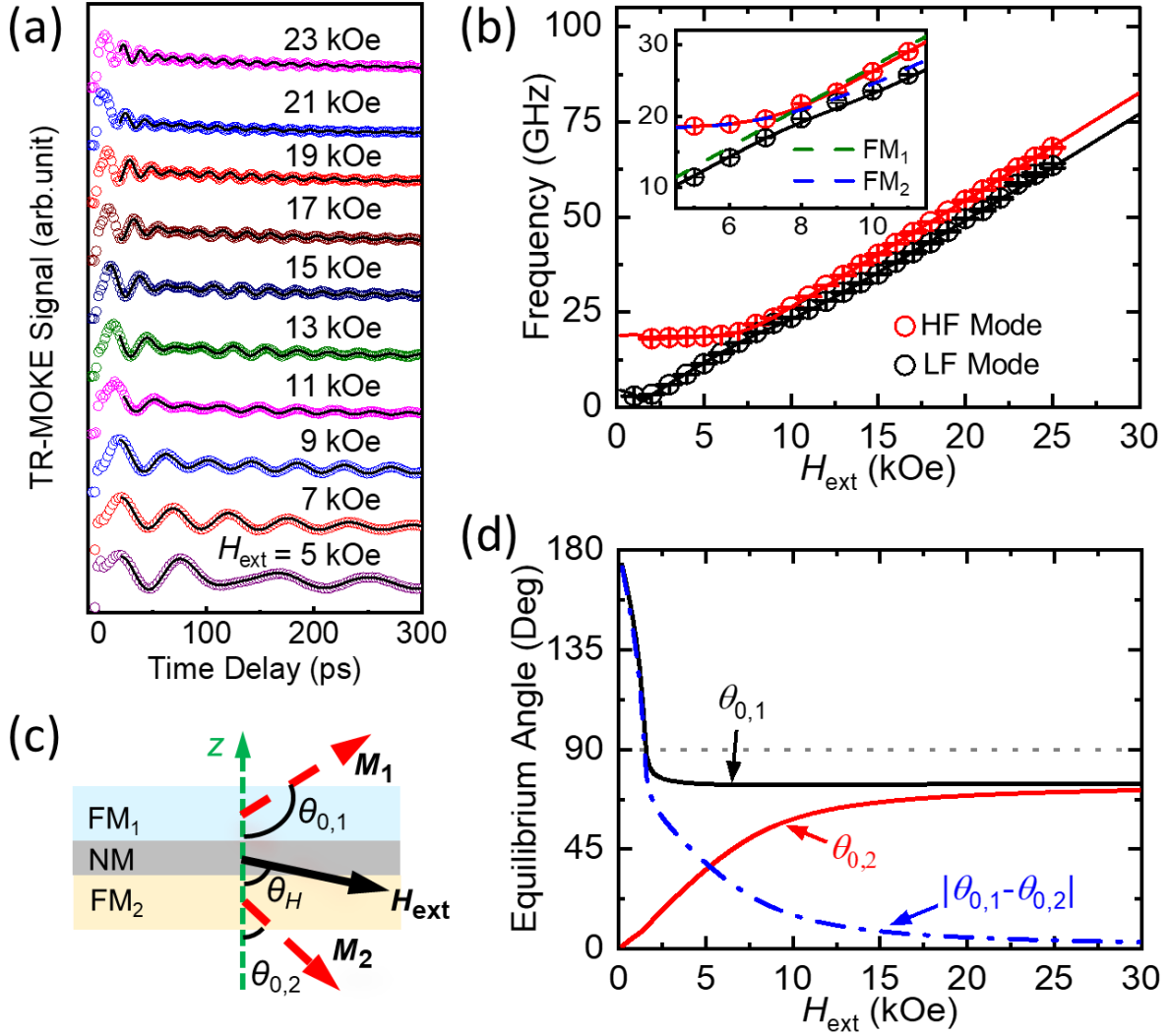

FIG. 2 (a) TR-MOKE signals under varying $H_{ext}$ when $\theta_H = 75°$ [as defined in panel (c)]. Circles are the experimental data and black lines are the fitting curves based on Eq. (6). (b) The precession frequencies of the HF and LF modes as functions of $H_{ext}$. Circles are experimental data and solid lines are fitting curves. The inset highlights the zoomed-in view of the field-dependent frequencies around 8 kOe, where the green dashed line and blue dashed line are the single-layer (SL) precession frequencies of $FM_1$ and $FM_2$ without interlayer exchange coupling. (c) Schematic illustration of the definition of the equilibrium polar angles ($\theta_{0,1}$ and $\theta_{0,2}$), and the direction of the external magnetic field ($\theta_H$). The illustration is equivalent to Fig. 1(b) due to symmetry. (d) $\theta_{0,1}$ and $\theta_{0,2}$ as functions of $H_{ext}$. The dash-dotted line plots the difference between the two equilibrium polar angles.

Based on the fitted stack properties ($H_{k,eff,1}$, $H_{k,eff,2}$, $J_1$, and $J_2$), the equilibrium magnetization directions in the two layers can be calculated. For SAFs with weak IEC compared with uniaxial PMA, the azimuthal angles of the magnetization in two FM layers are always the same as that of the external field at equilibrium status. Therefore, two polar angles will be sufficient to describe the equilibrium magnetization configuration. Figure 2(c) illustrates the definition of the equilibrium polar angles of two FM layers ($\theta_{0,1}$, $\theta_{0,2}$) and the external field ($\theta_H$). The values of $\theta_{0,1}$, $\theta_{0,2}$, and the difference between these two polar angles as functions of $H_{ext}$ are shown in Fig. 2(d). When $H_{ext}$ is low (< 1.6 kOe), magnetic anisotropy and antiferromagnetic coupling are dominant and $|\theta_{0,1} - \theta_{0,2}|$ is larger than 90°. As $H_{ext}$ increases, both $\theta_{0,1}$ and $\theta_{0,2}$ approach $\theta_H$. When $H_{ext}$ is high (> 15 kOe), the Zeeman energy becomes dominant and both $M_1$ and $M_2$ are almost aligned with $H_{ext}$.

## 3.2 Cone angle, direction, and phase of magnetization precession revealed by modeling

Besides the equilibrium configuration, using sample properties extracted from Fig. 2(b) as input parameters, the LLG-based modeling (described in section 2.2) also provides information on the cone angle, direction, and phase of magnetization precession for each mode (Fig. 1). The discussion in this section is limited to the case without damping and mutual spin pumping. They will be considered in Note 4 of the SM [35], sections 3.3, and 3.4. The calculation results are shown in Fig. 3, which are categorized into three regions. At high external fields ($H_{ext}$ > 1.6 kOe, regions 2 and 3), both FM layers precess CCW [$\text{Arg}\left(C_{\theta,i}/C_{\varphi,i}\right) = 90°$], and the polar angles of magnetization in two layers are in-phase [$\text{Arg}\left(C_{\theta,2}/C_{\theta,1}\right) = 0°$] for the HF mode and out-of-phase [$\text{Arg}\left(C_{\theta,2}/C_{\theta,1}\right) = 180°$] for the LF mode. This is the reason for the HF mode (LF mode) also being called the acoustic mode (optical mode) in the literature [23]. The criterion to differentiate region 2 from region 3 is the FM layer that dominates a given precessional mode (*i.e.*, the layer with larger precession cone angles). In region 2 (1.6 kOe < $H_{ext}$ < 8 kOe), the HF mode is dominated by $FM_2$ because $FM_2$ has larger cone angles than $FM_1$. This is reasonable since the higher precession frequency is closer to the natural frequency of $FM_2$ [see Fig. 2(b)] in region 2. Similarly, in region 3, the HF mode is dominated by $FM_1$ with larger precession cone angles.

When $H_{ext}$ is low (region 1), the angle between two magnetizations is larger than 90° [Fig. 2(d)] owing to the more dominant AF-exchange-coupling energy as compared with the Zeeman energy. In this region, magnetization dynamics exhibits some unique features. Firstly, CW [$\text{Arg}(C_{\theta,i}/C_{\varphi,i}) = -90°$] precession emerges: for each mode, the dominant layer precesses CCW ($FM_2$ for the HF mode and $FM_1$ for the LF mode) and the subservient layer precesses CW ($FM_1$ for the HF mode and $FM_2$ for the LF mode). This is because the effective field for the subservient layer [*e.g.*, $\mathbf{H}_{eff,1}$ for the HF mode, see Eq. (2)] precesses CW owing to the CCW precession of the dominant layer when $|\theta_{0,1} - \theta_{0,2}| > 90°$ [Fig. 2(d)]. In other words, a low $H_{ext}$ that makes $|\theta_{0,1} - \theta_{0,2}| > 90°$ is a necessary condition for the CW precession. However, it is not a sufficient condition. In general, certain degrees of symmetry breaking ($H_{k,eff,1} \neq H_{k,eff,2}$ or the field is tilted away from the direction normal to the easy axis) are also needed to generate CW precession. For example, for symmetric antiferromagnets ($H_{k,eff,1} = H_{k,eff,2}$) under fields perpendicular to the easy axis, CW precession does not appear even at low fields (Fig. 2(a) in Ref. [52]). See Note 5 of the SM [35] for more details. Secondly, as shown in Fig. 3, the precession motions in two FM layers are always in-phase for both HF and LF modes; thus, there is no longer a clear differentiation between "acoustic mode" and "optical mode". Instead, the two modes can be differentiated as "right-handed" and "left-handed" based on the chirality [53]. Here, we define the chirality with respect to a reference direction taken as the projection of $\mathbf{H}_{ext}$ or $\mathbf{M}_2$ (magnetization direction of the layer with a higher $H_{k,eff}$ on the easy axis [-$z$ direction in Fig. 3(c)]. Lastly, the shape of the precession cone also varies in different regions. $\Delta\theta_i$ and $\Delta\varphi_i$ are almost the same for both modes in region 3, indicating the precession trajectories are nearly circular. While in regions 1 and 2, $\Delta\theta_i$ and $\Delta\varphi_i$ are not always equal, suggesting the precession trajectories may have high ellipticities.

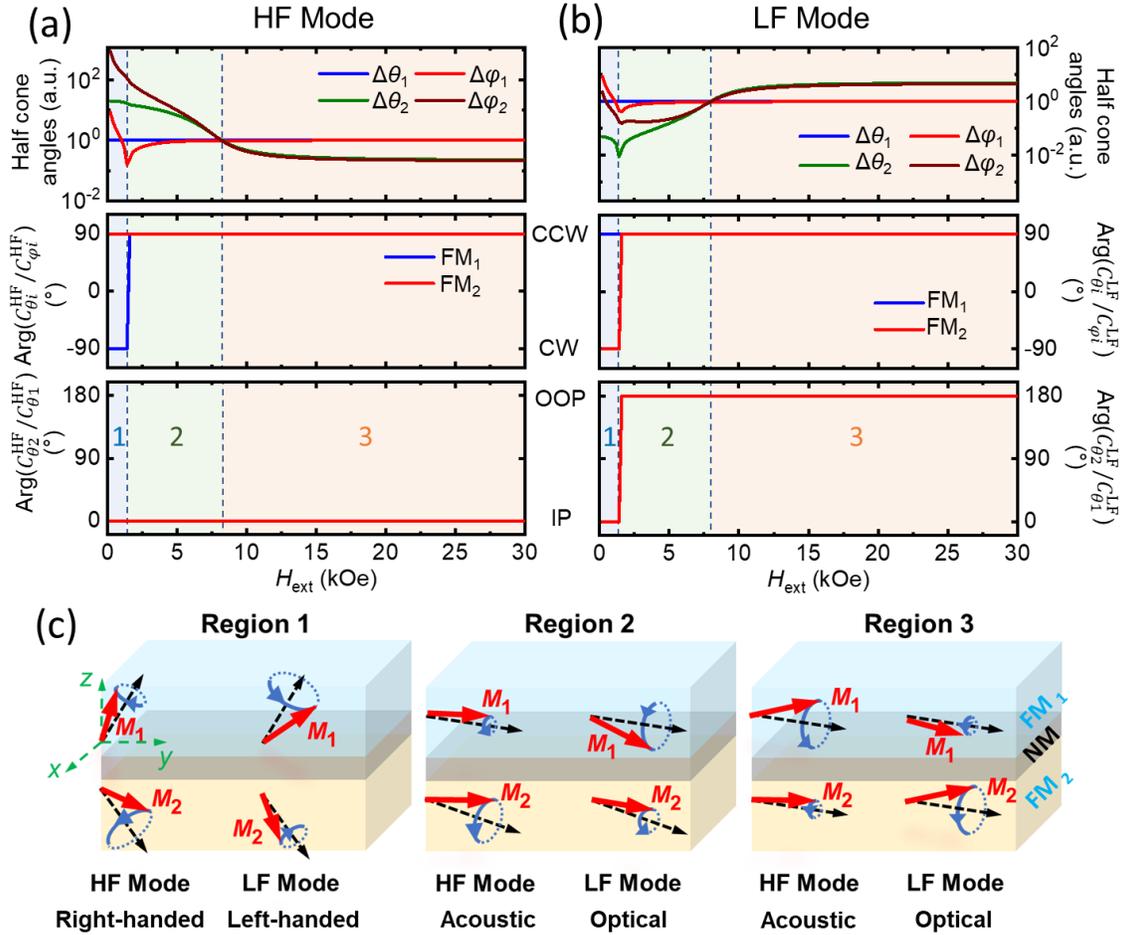

FIG. 3 The calculated half cone angle, direction, and phase of magnetization precession for (a) the HF mode and (b) the LF mode. In the top row, four curves represent the polar and azimuthal half cone angles of precession in two FM layers. All half cone angles are normalized with respect to $\Delta\theta_1$. The middle row shows the value of $\mathrm{Arg}\left(C_{\theta,i}/C_{\varphi,i}\right)$ under different $H_{ext}$. A value of 90° (−90°) represents CCW (CW) precession. The bottom row is the phase difference of the polar angles in two layers. A value of 0° (180°) corresponds to the polar angles of the magnetization in two layers are IP (OOP) during precession. Dashed lines correspond to the reference case where damping is zero in both layers. (c) Schematic illustrations of the cone angle, direction, and phase of magnetization precession for the HF and LF modes in different regions, and their corresponding characteristics regarding chirality and phase difference.

## 3.3    Amplitude and phase of TR-MOKE signals

Actual magnetization dynamics is resolvable as a linear combination of the two eigenmodes (the HF and the LF modes). By taking into account the initial conditions (*i.e.*, laser excitation, see Note 1 of the SM [35]), we can determine the amplitude and phase of the two modes in TR-MOKE signals. Figure 4(a) summarizes the amplitudes of both HF and LF modes [$C^{HF}$ and $C^{LF}$ in Eq. (3)] under different $H_{ext}$. Noted that the *y*-axis represents Kerr angle ($\theta_K$) instead of the cone angle of precession. The LF mode has a local minimum near 8 kOe, where the two FM layers have similar precession cone angles but opposite phases for the LF mode [Fig. 3(b)]. The amplitudes of both modes decrease with $H_{ext}$ in the high-field region. This is similar to the single-layer case, where the amplitudes of TR-MOKE signals decrease with $H_{ext}$ because the decrease in $H_{k,eff}$ induced by laser heating is not able to significantly alternate the equilibrium magnetization direction when the Zeeman energy dominates [51]. The LF mode also has an amplitude peak at low fields ($H_{ext} < 3$ kOe), where the dominant layer of FM$_1$ changes its equilibrium direction dramatically with $H_{ext}$ (from ~75° to 170°) as shown in Fig. 2(d).

To directly compare the amplitudes of TR-MOKE signals and the LLG-based calculations, the weighting factor *w* and the initial conditions are needed. The initial conditions are determined by $H'_{k,eff,1}$, $H'_{k,eff,2}$, and $J'_1$, representing the instantaneous effective anisotropy fields and IEC strength upon laser heating. These instantaneous properties are different from their corresponding room-temperature values ($H_{k,eff,1}$, $H_{k,eff,2}$, and $J_1$). The accurate determination of $H'_{k,eff,1}$, $H'_{k,eff,2}$, and $J'_1$ demands the modeling of the laser heating process as well as the temperature dependence of stack properties, which are challenging. Here, we treat these three variables as adjustable parameters and determine their values by fitting the field-dependent amplitudes of TR-MOKE signals, which yields $H'_{k,eff,1}/H_{k,eff,1} = 0.90 \pm 0.01$, $H'_{k,eff,2}/H_{k,eff,2} = 0.95 \pm 0.01$, and $J'_1/J_1 = 0.83 \pm 0.01$. It is apparent that the field dependence of TR-MOKE signal amplitude is in excellent agreement with the theoretical modeling, as shown in Fig. 4(a).

Figure 4(b) shows the calculated half polar cone angles for each mode in each FM layer. In TR-MOKE signals, the optical mode (the LF mode in regions 2 and 3) tends to be partially canceled out because the two layers precess out-of-phase. Therefore, compared with Fig. 4(a), the information in Fig. 4(b) better reflects the actual intensity of both modes in $FM_1$ and $FM_2$. In Fig. 4(b), the precession cone angles of both modes in $FM_1$ ($\Delta\theta_1^{HF}, \Delta\theta_1^{LF}$) have local maxima at the anti-crossing field ($H_{ext} \approx 8$ kOe). On the contrary, $\Delta\theta_2^{LF}$ and $\Delta\theta_2^{HF}$ of $FM_2$ have their maxima either above or below the anti-crossing field. This is because $FM_2$ has larger precession amplitudes (cone angles) than $FM_1$ at the anti-crossing field if there is no IEC [the dotted lines of $FM_1$ (SL) and $FM_2$ (SL) in Fig. 4(b)]. With IEC, $FM_2$ with larger cone angles can drive the precession motion in $FM_1$ significantly near the anti-crossing field, where IEC is effective. Subsequently, the precession amplitudes of $FM_1$ exhibit local maxima as its cone angle peaks at the anti-crossing field [solid lines in Fig. 4(b)]. Also, compared with the uncoupled case [$FM_1$ (SL) in Fig. 4(b)], $FM_1$ in the SAF structure has a much larger cone angle at the boundary between regions 1 and 2 ($H_{ext} \approx 1.6$ kOe). This corresponds to the case where $FM_1$ fast switching is driven by $H_{ext}$, as shown in Fig. 2(d). The energy valley of $FM_1$ created by IEC and uniaxial anisotropy is canceled out by $H_{ext}$. As a result, any perturbation in $H_{k,eff,1}$ or IEC can induce a large change in $\theta_1$.

Besides amplitude, the phase of TR-MOKE signals [$\beta^{HF}$ and $\beta^{LF}$ in Eq. (6)] also provides important information about the magnetization dynamics in SAF [Fig. 4(c)]. In Fig. 4(c), the phase of the HF mode stays constant around $\pi$. However, the LF mode goes through a $\pi$-phase shift at the transition from region 2 to region 3. This phase shift can be explained by the change of the dominant layer from region 2 to region 3 for the LF mode [Fig. 3(c)]. As illustrated in Fig. 4(d), the LF mode (optical mode in regions 2 and 3) has opposite phases in $FM_1$ (~0°) and $FM_2$ (~180°). Considering the two FM layers have comparable optical contributions to TR-MOKE signals ($w \approx$ 0.5), TR-MOKE signals will reflect the phase of the dominant layer for each mode. In region 3, $FM_2$ has larger precession cone angles than $FM_1$ for the LF mode; therefore, LF TR-MOKE signals have the same phase as $FM_2$ (~180°). However, in region 2, the dominant layer shifts from $FM_2$ to $FM_1$ for the LF mode. Hence, the phase of LF TR-MOKE signals also changes by ~180° to be consistent with the phase of $FM_1$ (~0°). As for the HF mode, since the two layers always have almost the same phase (~180°), the change of the dominant layer does not cause a shift in the phase of TR-MOKE signals.

By comparing Fig. 4(d) and Fig. 3(a-b), one can notice that the phase difference between two FM layers could deviate from 0° or 180° when damping and mutual spin pumping is considered [Fig. 4(d)]. The deviation of phase allows energy to be transferred from one FM layer to the other during precession via exchange coupling [54]. In our sample system, $FM_2$ has a higher damping constant ($\alpha_1 = 0.020$ and $\alpha_2 = 0.060$); therefore, the net transfer of energy is from $FM_1$ to $FM_2$. More details can be found in Note 4 of the SM [35], which shows the phase of TR-MOKE signals is affected by Gilbert damping in both layers and the mutual spin pumping. By fitting the phase [Fig. 4(c)] and the damping [Fig. 5(a)] of TR-MOKE signals simultaneously, we obtained $\alpha_{sp,12}$ = 0.010 ± 0.004, $\alpha_{sp,21} = 0.007^{+0.009}_{-0.007}$, $\alpha_1 = 0.020 \pm 0.002$, and $\alpha_2 = 0.060 \pm 0.008$. Nonreciprocal spin pumping damping ($\alpha_{sp,12} \neq \alpha_{sp,21}$) has been reported in asymmetric $FM_1/NM/FM_2$ trilayers and attributed to the different spin-mixing conductance ($g_i^{\uparrow\downarrow}$) at the two FM/NM interfaces [27], following $\alpha_{sp,ij} = g_i \mu_B g_j^{\uparrow\downarrow}/(8\pi M_{s,i} d_i)$, with $g_i$ the $g$-factor of the $i$-th layer and $\mu_B$ the Bohr magneton [55]. The above equation neglects the spin-flip scattering in NM and assumes that the spin accumulation in the NM spacer equally flows back to FM$_1$ and FM$_2$ [37]. However, the uncertainties of our $\alpha_{\text{sp},ij}$ are too high to justify the nonreciprocity of $\alpha_{\text{sp},ij}$ (see Note 3 of the SM [35] for detailed uncertainty analyses). In fact, if the spin backflow to FM$_i$ is proportional to $g_i^{\uparrow\downarrow}$, then $\alpha_{\text{sp},ij} = g_i\mu_B g_i^{\uparrow\downarrow} g_j^{\uparrow\downarrow} / \left[4\pi M_{\text{s},i} d_i \left(g_i^{\uparrow\downarrow} + g_j^{\uparrow\downarrow}\right)\right]$ (Eq. 1.14 in Ref.[56]). In this case, the different spin-mixing conductance at two FM/NM interfaces ($g_1^{\uparrow\downarrow} \neq g_2^{\uparrow\downarrow}$) will not lead to nonreciprocal $\alpha_{\text{sp},ij}$. Although differences in $g_i$ and magnetic moment per area ($M_{\text{s},i} d_i$) can potentially lead to nonreciprocal $\alpha_{\text{sp},ij}$, the values of $g_i$ and $M_{\text{s},i} d_i$ for the two FM layers are expected to be similar (the net magnetization of SAF is zero without external fields). Therefore, nearly reciprocal $\alpha_{\text{sp},ij}$ are plausible for our SAF stack. Assuming $g_i^{\uparrow\downarrow}$ values are similar at the two FM/NM interfaces ($g_1^{\uparrow\downarrow} \approx g_2^{\uparrow\downarrow} = g^{\uparrow\downarrow}$), this yields $g^{\uparrow\downarrow} = 8\pi M_{\text{s},i} d_i \alpha_{\text{sp},ij} / (g_i \mu_B) = 1.2 \sim 1.7 \times 10^{15}$ cm$^{-2}$. $g^{\uparrow\downarrow}$ can also be estimated from the free electron density per spin ($n$) in the NM layer: $g^{\uparrow\downarrow} \approx 1.2n^{2/3}$ [57]. With $n = 5.2 \times 10^{28}$ m$^{-3}$ for Ru [58] (the value of $n$ is similar for Ta [59]), $g^{\uparrow\downarrow}$ is estimated to be $1.7 \times 10^{15}$ cm$^{-2}$, the same order as the $g^{\uparrow\downarrow}$ value from TR-MOKE measurements, which justifies the $\alpha_{\text{sp},ij}$ values derived from TR-MOKE are within a reasonable range. The values of $\alpha_1$ and $\alpha_2$ will be discussed in section 3.4.

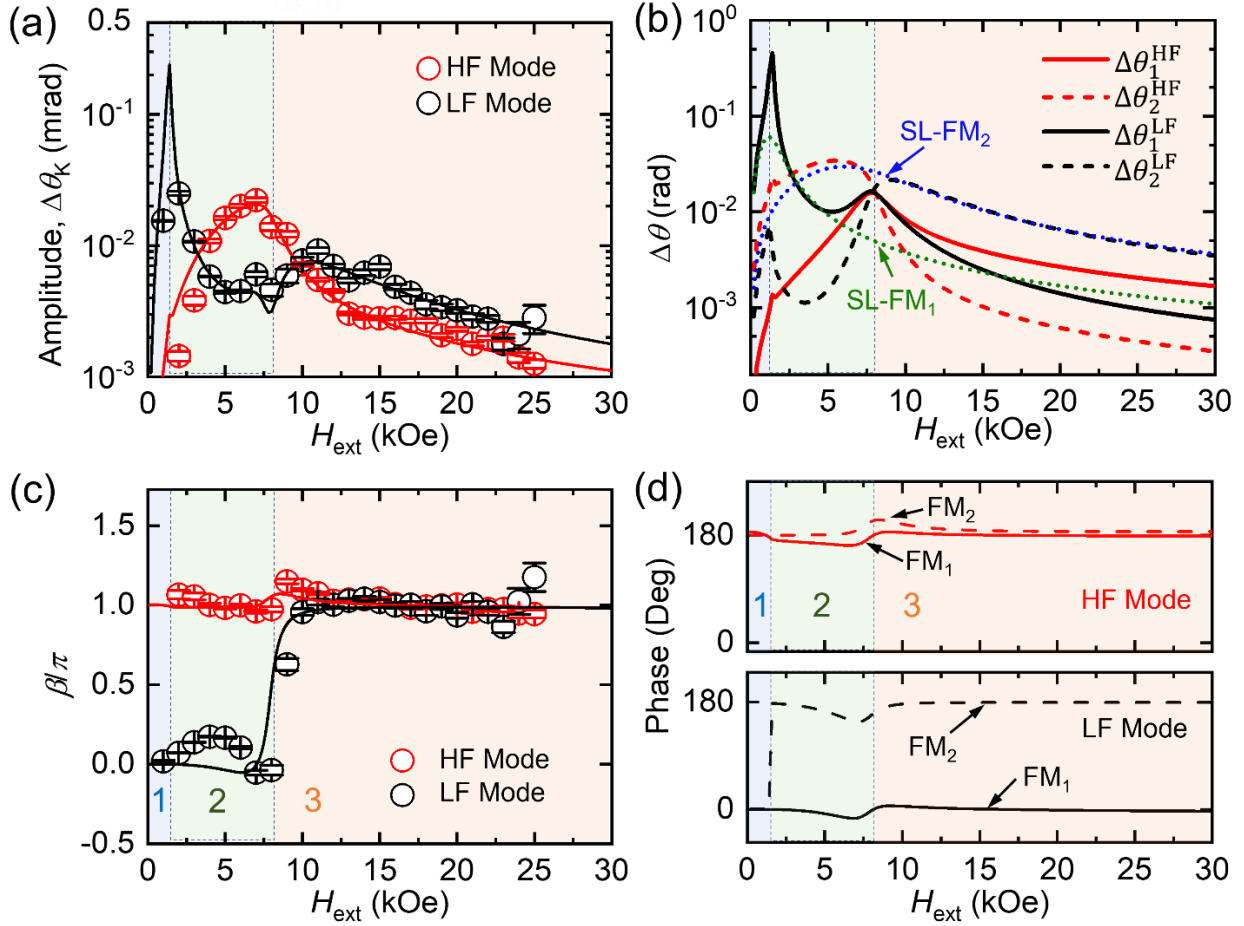

FIG. 4 (a) Amplitudes of TR-MOKE signals as functions of $H_{ext}$. The circles and curves represent experimental data and modeling fitting, respectively. (b) The calculated precession half cone angles at different $H_{ext}$. Red curves and black curves represent the cone angles of the HF mode and the LF mode in $FM_1$ (solid lines) and $FM_2$ (dashed lines). Dotted lines are the precession cone angles of single-layer (SL) $FM_1$ and $FM_2$ without IEC. (c) Phases of TR-MOKE signals at varying $H_{ext}$. Circles and curves are experimental data and modeling fitting ($\alpha_{sp,12} = 0.010$, $\alpha_{sp,21} = 0.007$, $\alpha_1 = 0.020$, $\alpha_2 = 0.060$). (d) Simulated precession phase of the HF mode (red curves) and the LF mode (black curves) in $FM_1$ (solid lines) and $FM_2$ (dashed lines).

### 3.4 Magnetic damping of the HF and LF precession modes

In addition to the amplitude and phase of TR-MOKE signals for the p-SAF stack, the model analyses also provide a better understanding of magnetic damping. Figure 5(a) shows the effective damping constant ($\alpha_{eff} = 1/2\pi f \tau$) measured at different $H_{ext}$ (symbols), in comparison with modeling fitting (solid lines). The general $H_{\text{ext}}$ dependence of $\alpha_{\text{eff}}$ can be well captured by the model. The fitted Gilbert damping, $\alpha_1 = 0.020 \pm 0.002$ and $\alpha_2 = 0.060 \pm 0.008$ are close to the Gilbert damping of Ta/CoFeB(1 nm)/MgO thin films (~0.017) [41,60] and Co/Pd multilayers with a similar $t_{\text{Co}}/t_{\text{Pd}}$ ratio (~0.085) [61]. Other fitted parameters are $\Delta H_{\text{k,eff,1}} = 0.26 \pm 0.02$ kOe, $\Delta H_{\text{k,eff,2}} = 1.42 \pm 0.18$ kOe, $\alpha_{12}^{\text{sp}} = 0.010 \pm 0.004$ $\alpha_{21}^{\text{sp}} = 0.007_{-0.007}^{+0.009}$. $\Delta J_1$ and $\Delta J_2$ are set to be zero, as explained in Sec. 2.2. More details regarding the values and determination methods of all parameters involved in our data reduction are provided in Note 3 of the SM [35]. Dashed lines show the calculated $\alpha_{\text{eff}}$ without inhomogeneous broadening. At high $H_{\text{ext}}$, the difference between the solid lines and dashed lines approaches zero because the inhomogeneous broadening is suppressed. At low $H_{\text{ext}}$, the solid lines are significantly higher than the dashed lines, indicating substantial inhomogeneous broadening contributions.

The effective damping shows interesting features near the anti-crossing field. As shown in Fig. 5(b), due to the effective coupling between two FM layers near the anti-crossing field, the hybridization of precession in two FM layers leads to a mix of damping with contributions from both layers. The effective damping of the $FM_1$-dominant mode reaches a maximum within the anti-crossing region ($7 \leq H_{\text{ext}} \leq 10$ kOe) and is higher than the single-layer (SL) $FM_1$ case. Similarly, the hybridized HF and LF modes at 8.5 kOe exhibit a lower $\alpha_{\text{eff}}$ (~0.073) compared to the SL $FM_2$ case. $\alpha_{\text{eff}}$ consists of contributions from Gilbert damping ($\alpha_i$), mutual spin pumping ($\alpha_{\text{sp},ij}, i \neq j$), and inhomogeneous broadening ($\Delta H_{\text{k,eff},i}$ and $\Delta J_i$). To better understand the mixing damping behavior, Fig. 5(c) shows $\alpha_{\text{eff}}$ after excluding the inhomogeneous contribution ($\alpha_{\text{eff}}^{\text{inhomo}}$). Compared to the SL layer case (green and blue dashed lines), the HF and LF modes (red and black dashed lines) clearly suggest that IEC effectively mixes the damping in two layers around the anti-crossing field. Without the IEC, precession in $FM_2$ with a higher damping relaxes faster than that in FM$_1$. However, the IEC provides a channel to transfer energy from FM$_1$ to FM$_2$, such that the two layers have the same precession relaxation rate for a given mode. Near the anti-crossing field, two layers have comparable precession cone angles; therefore, the damping values of the hybridized modes are roughly the average of two FM layers. In addition to the static IEC, dynamic spin pumping can also modify the damping of individual modes. The black and red solid lines represent the cases with mutual spin pumping ($\alpha_{\mathrm{sp,12}} = 0.01$ and $\alpha_{\mathrm{sp,21}} = 0.007$). Generally, in regions 2 & 3, mutual spin pumping reduces the damping of the HF mode and increases the damping of the LF mode because the HF (LF) mode is near in-phase (out-of-phase). Overall, the static IEC still plays the essential role for the damping mix near the anti-crossing field.

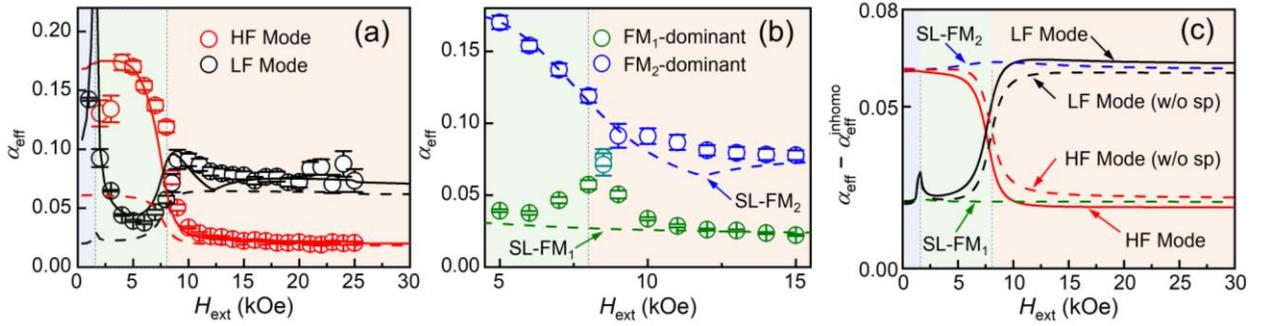

FIG. 5 (a) Effective damping constant under varying $H_{\mathrm{ext}}$. Circles are experimental data. Solid lines are fitting curves based on Eqs. (4-5). Dashed lines denote $\alpha_{\mathrm{eff}}$ after the removal of inhomogeneous-broadening contribution. (b) A zoomed-in figure of panel (a) between 5 kOe and 15 kOe. Blue and green circles are measured effective damping of the mode dominated by FM$_1$ and FM$_2$, respectively. Blue and green dashed lines are the $\alpha_{\mathrm{eff}}$ of FM$_1$ and FM$_2$ single layer without IEC. (c) Effective damping after excluding the inhomogeneous contribution as a function of $H_{\mathrm{ext}}$. The HF mode (red curves) and the LF mode (black curves) are represented by solid (or dashed) curves when the mutual spin pumping terms ($\alpha_{\mathrm{sp,12}}$ and $\alpha_{\mathrm{sp,21}}$) are considered (or excluded). The dashed green and blue lines are the SL cases for FM$_1$ and FM$_2$, respectively.

## 4 CONCLUSION

We systematically investigated the magnetization dynamics excited by ultrafast laser pulses in an asymmetric p-SAF sample both theoretically and experimentally. We obtained detailed information regarding magnetization dynamics, including the cone angles, directions, and phases of spin precession in each layer under different $H_{ext}$. In particular, the dynamic features in the low-field region (region 1) exhibiting CW precession, were revealed. The resonance between the precession of two FM layers occurs at the boundary between regions 2 and 3, where an anti-crossing feature is present in the frequency $vs.$ $H_{ext}$ profile. The dominant FM layer for a given precession mode also switches from region 2 to region 3. The amplitude and phase of TR-MOKE signals are well captured by theoretical modeling. Importantly, we successfully quantified the individual contributions from various sources to the effective damping, which enables the determination of Gilbert damping for both FM layers. At low $H_{ext}$, the contribution of inhomogeneous broadening to the effective damping is significant. Near the anti-crossing field, the effective damping of two coupled modes contains substantial contributions from both FM layers owing to the strong hybridization via IEC. Although the analyses were made for an asymmetric SAF sample, this approach can be directly applied to study magnetization dynamics and magnetic properties of general complex material systems with coupled multilayers, and thus benefits the design and optimization of spintronic materials via structural engineering.

**Acknowledgements**

This work is primarily supported by the National Science Foundation (NSF, CBET- 2226579). D.L.Z gratefully acknowledges the funding support from the ERI program (FRANC) "Advanced MTJs for computation in and near random access memory" by DARPA, and ASCENT, one of six centers in JUMP (a Semiconductor Research Corporation program, sponsored by MARCO and DARPA). J.P.W and X.J.W also appreciate the partial support from the UMN MRSEC Seed program (NSF, DMR-2011401). D.B.H. would like to thank the support from the UMN 2022-2023 Doctoral Dissertation Fellowship. The authors appreciated the valuable discussion with Prof. Paul Crowell.

# Supplemental Material for

# Magnetization Dynamics in Synthetic Antiferromagnets with Perpendicular Magnetic Anisotropy

Dingbin Huang[1,*], Delin Zhang[2], Yun Kim[1], Jian-Ping Wang[2], and Xiaojia Wang[1,*]

[1]Department of Mechanical Engineering, University of Minnesota, Minneapolis, MN 55455, USA
[2]Department of Electrical and Computer Engineering, University of Minnesota, Minneapolis, MN 55455, USA

**Supplemental Note 1: Analyses of the magnetization precession in each Ferromagnetic (FM) layer**

For the convenience of derivation, $\mathbf{m}_i$ is represented in the spherical coordinates with the polar angle $\theta_i$ and the azimuthal angle $\varphi_i$, as shown in Fig. 1(b):

$$\mathbf{m}_i = (\sin\,\theta_i \cos\,\varphi_i, \sin\,\theta_i \sin\,\varphi_i, \cos\,\theta_i) \tag{S1}$$

Accordingly, the expression of Eq. (2) in the spherical coordinates is:

$$
\begin{cases}
\dot{\theta}_1 = \dfrac{-\gamma_1}{d_1 M_{s,1}\sin\theta_1}\dfrac{\partial F}{\partial \varphi_1} - \alpha_1\sin\theta_1\dot{\varphi}_1 + \alpha_{sp,12}\sin\theta_2\cos(\theta_2-\theta_1)\dot{\varphi}_2 \\[2mm]
\dot{\varphi}_1 = \dfrac{\gamma_1}{d_1 M_{s,1}\sin\theta_1}\dfrac{\partial F}{\partial \theta_1} + \dfrac{\alpha_1}{\sin\theta_1}\dot{\theta}_1 - \dfrac{\alpha_{sp,12}}{\sin\theta_1}\dot{\theta}_2 \\[2mm]
\dot{\theta}_2 = \dfrac{-\gamma_2}{d_2 M_{s,2}\sin\theta_2}\dfrac{\partial F}{\partial \varphi_2} - \alpha_2\sin\theta_2\dot{\varphi}_2 + \alpha_{sp,21}\sin\theta_1\cos(\theta_1-\theta_2)\dot{\varphi}_1 \\[2mm]
\dot{\varphi}_2 = \dfrac{\gamma_2}{d_2 M_{s,2}\sin\theta_2}\dfrac{\partial F}{\partial \theta_2} + \dfrac{\alpha_2}{\sin\theta_2}\dot{\theta}_2 - \dfrac{\alpha_{sp,21}}{\sin\theta_2}\dot{\theta}_1
\end{cases}
\tag{S2}
$$

*Authors to whom correspondence should be addressed: huan1746@umn.edu and wang4940@umn.edu

where, a dot over variables represents a derivative with respect to time. When $\mathbf{M}_i$ precesses around its equilibrium direction:

$$\begin{cases} \theta_i = \theta_{0,i} + \Delta\theta_i \\ \varphi_i = \varphi_{0,i} + \Delta\varphi_i \end{cases} \tag{S3}$$

with $\Delta\theta_i$ and $\Delta\varphi_i$ representing the deviation angles of $\mathbf{M}_i$ from its equilibrium direction along the polar and azimuthal directions. Assuming the deviation is small, under the first-order approximation, the first-order partial derivative of $F$ in Eq. (S2) can be expanded as:

$$\begin{cases} \dfrac{\partial F}{\partial \theta_i} \approx \dfrac{\partial^2 F}{\partial \theta_i^2}\Delta\theta_i + \dfrac{\partial^2 F}{\partial \varphi_i\,\partial \theta_i}\Delta\varphi_i + \dfrac{\partial^2 F}{\partial \theta_j\,\partial \theta_i}\Delta\theta_j + \dfrac{\partial^2 F}{\partial \varphi_j\,\partial \theta_i}\Delta\varphi_j \\[2mm] \dfrac{\partial F}{\partial \varphi_i} \approx \dfrac{\partial^2 F}{\partial \theta_i\,\partial \varphi_i}\Delta\theta_i + \dfrac{\partial^2 F}{\partial \varphi_i^2}\Delta\varphi_i + \dfrac{\partial^2 F}{\partial \theta_j\,\partial \varphi_i}\Delta\theta_j + \dfrac{\partial^2 F}{\partial \varphi_j\,\partial \varphi_i}\Delta\varphi_j \end{cases} \tag{S4}$$

By substituting Eq. (S4), Equation (S2) is linearized as [1]:

$$\begin{bmatrix} \Delta\dot{\theta}_1 \\ \Delta\dot{\varphi}_1 \\ \Delta\dot{\theta}_2 \\ \Delta\dot{\varphi}_2 \end{bmatrix} = \mathbf{K} \begin{bmatrix} \Delta\theta_1 \\ \Delta\varphi_1 \\ \Delta\theta_2 \\ \Delta\varphi_2 \end{bmatrix} \tag{S5}$$

where, $\mathbf{K}$ is a 4×4 matrix, consisting of the properties of individual FM layers and the second-order derivatives of $F$ in terms of $\theta_1$, $\varphi_1$, $\theta_2$, and $\varphi_2$. Equation (S5) has four eigen-solutions, in the form of $C\exp(i\omega t)$, corresponding to four precession frequencies: $\pm\omega^{\mathrm{HF}}$ and $\pm\omega^{\mathrm{LF}}$. A pair of eigen-solutions with the same absolute precession frequency are physically equivalent. Therefore, only two eigen-solutions need to be considered:

$$\begin{cases} \Delta\theta_i = C_{\theta,i}^{\mathrm{HF}}\exp\left(i\omega^{\mathrm{HF}}t\right) \\ \Delta\varphi_i = C_{\varphi,i}^{\mathrm{HF}}\exp\left(i\omega^{\mathrm{HF}}t\right) \end{cases} \text{and} \begin{cases} \Delta\theta_i = C_{\theta,i}^{\mathrm{LF}}\exp\left(i\omega^{\mathrm{LF}}t\right) \\ \Delta\varphi_i = C_{\varphi,i}^{\mathrm{LF}}\exp\left(i\omega^{\mathrm{LF}}t\right) \end{cases} \tag{S6}$$

After rearrangement, the full solutions in the spherical coordinates are expressed as below (also Eq. (3) in the main paper).

$$\begin{bmatrix} \theta_1(t) \\ \varphi_1(t) \\ \theta_2(t) \\ \varphi_2(t) \end{bmatrix} = \begin{bmatrix} \theta_{0,1} \\ \varphi_{0,1} \\ \theta_{0,2} \\ \varphi_{0,2} \end{bmatrix} + \begin{bmatrix} \Delta\theta_1(t) \\ \Delta\varphi_1(t) \\ \Delta\theta_2(t) \\ \Delta\varphi_2(t) \end{bmatrix} = \begin{bmatrix} \theta_{0,1} \\ \varphi_{0,1} \\ \theta_{0,2} \\ \varphi_{0,2} \end{bmatrix} + \begin{bmatrix} C_{\theta,1}^{\mathrm{HF}} \\ C_{\varphi,1}^{\mathrm{HF}} \\ C_{\theta,2}^{\mathrm{HF}} \\ C_{\varphi,2}^{\mathrm{HF}} \end{bmatrix} \exp\left(i\omega^{\mathrm{HF}}t\right) + \begin{bmatrix} C_{\theta,1}^{\mathrm{LF}} \\ C_{\varphi,1}^{\mathrm{LF}} \\ C_{\theta,2}^{\mathrm{LF}} \\ C_{\varphi,2}^{\mathrm{LF}} \end{bmatrix} \exp\left(i\omega^{\mathrm{LF}}t\right) \qquad (S7)$$

The prefactors of these eigen-solutions provide information about magnetization dynamics of both the HF and LF modes. Directly from solving Eq. (S2), one can obtain the relative ratios of these prefactors, which are $\left[C_{\varphi_1}^{\mathrm{HF}}/C_{\theta_1}^{\mathrm{HF}}, C_{\theta_2}^{\mathrm{HF}}/C_{\theta_1}^{\mathrm{HF}}, C_{\varphi_2}^{\mathrm{HF}}/C_{\theta_1}^{\mathrm{HF}}\right]$ and $\left[C_{\varphi_1}^{\mathrm{LF}}/C_{\theta_1}^{\mathrm{LF}}, C_{\theta_2}^{\mathrm{LF}}/C_{\theta_1}^{\mathrm{LF}}, C_{\varphi_2}^{\mathrm{LF}}/C_{\theta_1}^{\mathrm{LF}}\right]$. These ratios provide precession information of each mode, as presented in Fig. 3.

Obtaining the absolute values of $\left[C_{\theta,1}, C_{\varphi,1}, C_{\theta,2}, C_{\varphi,2}\right]^T$ for each mode requires the initial conditions of precession, which is necessary for fitting the actual precession amplitudes in TR-MOKE signals. In TR-MOKE measurements, magnetization precession is initiated by laser heating, which reduces the magnetic anisotropy of each FM layer and the interlayer exchange coupling strength between two FM layers [2]. Considering the laser heating process is ultrafast compared with magnetization precession while the following cooling due to heat dissipation is much slower than magnetization dynamics, we approximately model the temporal profiles of effective anisotropy fields and exchange coupling as step functions. Owing to the sudden change in magnetic properties induced by laser heating, magnetization in each layer will establish a new equilibrium direction $\left(\theta'_{0,i}, \varphi'_{0,i}\right)$. In other words, $\mathbf{M}_i$ deviates from its new equilibrium direction by $\Delta\theta_i = \theta_{0,i} - \theta'_{0,i}$, $\Delta\varphi_i = \varphi_{0,i} - \varphi'_{0,i}$. Substituting $t = 0$ to Eq. (S7), one can get the initial conditions for magnetization dynamics:

$$\Delta\theta_i(t=0) = C_{\theta,i}^{\mathrm{HF}} + C_{\theta,i}^{\mathrm{LF}} = \theta_{0,i} - \theta'_{0,i}$$
$$\Delta\varphi_i(t=0) = C_{\varphi,i}^{\mathrm{HF}} + C_{\varphi,i}^{\mathrm{LF}} = \varphi_{0,i} - \varphi'_{0,i} = 0$$
$$(S8)$$

Once the initial conditions are set, the absolute values of all prefactors can be obtained.

**Supplemental Note 2: Estimation of each layer's contribution to total TR-MOKE signals**

The contribution from each FM layer is estimated by static MOKE measurement. According to Ref. [3], the result from this method matches well with that from the optical calculation. The sample is perpendicularly saturated before the static MOKE measurement. Then the out-of-plane $M$-$H_{ext}$ loop (Fig. S1) is measured by static MOKE. As shown in the figure, two different antiferromagnetic (AF) configurations have different normalized MOKE signals, indicating the different contributions to the total signals by two layers. The weighting factor is calculated by:

$$-w + (1 - w) = 0.085 \tag{S9}$$

which gives $w = 0.457$. Considering the relatively small layer thicknesses [$FM_1$: CoFeB(1), spacer: Ru(0.6)/Ta(0.3), and $FM_2$: Co(0.4)/Pd(0.7)/Co(0.4)], it is reasonable that $FM_1$ and $FM_2$ make comparable contributions to the total TR-MOKE signals (*i.e.*, $w \approx 0.5$).

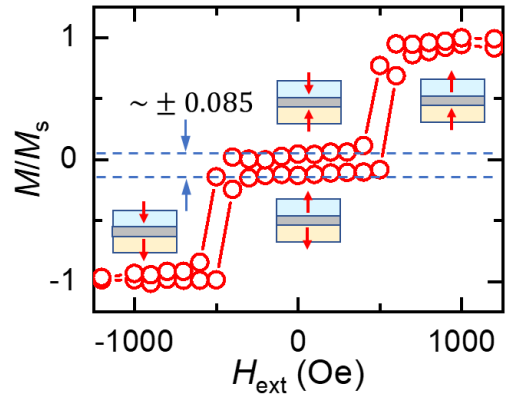

FIG. S1 Static MOKE hysteresis loop. Magnetic fields are applied along the out-of-plane direction.

**Supplemental Note 3: Summary of the parameters and uncertainties for data reduction**

Given that a number of variables are involved in the analysis, TABLE SI summarizes the major variables discussed in the manuscript, along with their values and determination methods.

TABLE SI. Summary of the values and determination methods of parameters used in the data reduction. The reported uncertainties are one-sigma uncertainties from the mathematical model fitting to the TR-MOKE measurement data.

| Parameters | Values | Determination Methods |
|---|---|---|
| $H_f$ | ~500 Oe | VSM |
| $M_{s,1}$ | 1240 emu cm$^{-3}$ | VSM |
| $M_{s,2}$ | 827 emu cm$^{-3}$ | VSM |
| $d_1$ | 1 nm | Sample structure |
| $d_2$ | 1.5 nm | Sample structure |
| $H_{k,eff,1}$ | $1.23 \pm 0.28$ kOe | Fitted from $f$ vs. $H_{ext}$ [Fig. 2(b)] |
| $H_{k,eff,2}$ | $6.18 \pm 0.13$ kOe | Fitted from $f$ vs. $H_{ext}$ [Fig. 2(b)] |
| $\gamma_1$ | $17.79 \pm 0.04$ rad ns$^{-1}$ kOe$^{-1}$ | Fitted from $f$ vs. $H_{ext}$ [Fig. 2(b)] |
| $\gamma_2$ | $17.85 \pm 0.04$ rad ns$^{-1}$ kOe$^{-1}$ | Fitted from $f$ vs. $H_{ext}$ [Fig. 2(b)] |
| $J_1$ | $-0.050 \pm 0.020$ erg cm$^{-2}$ | Fitted from $f$ vs. $H_{ext}$ [Fig. 2(b)] |
| $J_2$ | 0 | Fitted from $f$ vs. $H_{ext}$ [Fig. 2(b)] |
| $w$ | 0.457 | Static MOKE |
| $H'_{k,eff,1}/H_{k,eff,1}$ | $0.90 \pm 0.01$ | Fitted from Amp vs. $H_{ext}$ [Fig. 4(a)] |
| $H'_{k,eff,2}/H_{k,eff,2}$ | $0.95 \pm 0.01$ | Fitted from Amp vs. $H_{ext}$ [Fig. 4(a)] |
| $J'_1/J_1$ | $0.83 \pm 0.01$ | Fitted from Amp vs. $H_{ext}$ [Fig. 4(a)] |
| $\alpha_1$ | $0.020 \pm 0.002$ | Fitted from $\alpha_{eff}$ vs. $H_{ext}$ [Fig. 5(a)] and $\beta$ vs. $H_{ext}$ [Fig. 4(c)] |
| $\alpha_2$ | $0.060 \pm 0.008$ | Fitted from $\alpha_{eff}$ vs. $H_{ext}$ [Fig. 5(a)] and $\beta$ vs. $H_{ext}$ [Fig. 4(c)] |
| $\Delta H_{k,eff,1}$ | $0.26 \pm 0.02$ kOe | Fitted from $\alpha_{eff}$ vs. $H_{ext}$ [Fig. 5(a)] |
| $\Delta H_{k,eff,2}$ | $1.42 \pm 0.18$ kOe | Fitted from $\alpha_{eff}$ vs. $H_{ext}$ [Fig. 5(a)] |
| $\alpha_{sp,12}$ | $0.010 \pm 0.004$ | Fitted from $\alpha_{eff}$ vs. $H_{ext}$ [Fig. 5(a)] and $\beta$ vs. $H_{ext}$ [Fig. 4(c)] |
| $\alpha_{sp,21}$ | $0.007^{+0.009}_{-0.007}$ | Fitted from $\alpha_{eff}$ vs. $H_{ext}$ [Fig. 5(a)] and $\beta$ vs. $H_{ext}$ [Fig. 4(c)] |

**Supplemental Note 4: Impacts of $\alpha_1$, $\alpha_2$, and mutual spin pumping on the phase**

Without damping, the phase difference in the precession polar angles of two FM layers [Arg($C_{\theta_2}/C_{\theta_1}$)] is always 0° or 180°, as shown in Fig. 3 of the main article. However, this does not necessarily hold if either the damping or mutual spin pumping is considered. The changes in the phase difference due to damping are depicted in Fig. S2. When $\alpha_1 = \alpha_2$, the phase difference between two layers stays at 0° or 180° [Fig. S2(a)], identical to the lossless case ($\alpha_1 = \alpha_2 = 0$) in Fig. 3. As a result, the initial phase of TR-MOKE signals ($\beta$) also stays at 0° or 180° [Fig. S2(b)]. However, when $\alpha_1 \neq \alpha_2$, Arg($C_{\theta_2}/C_{\theta_1}$) deviates from 0° or 180° especially at high fields ($H_{\text{ext}} >$ 5 kOe) [Fig. S2(c,e)]. The layer with a higher damping [FM$_1$ in (c) or FM$_2$ in (e)] tends to have a more advanced phase at high fields (regions 2 and 3). For example, in Fig. S2(e), 0° < Arg($C_{\theta_2}/C_{\theta_1}$) < 180° for both HF and LF modes in regions 2 and 3. The deviation from the perfect in-phase (0°) or out-of-phase (180°) condition allows the IEC to transfer energy from the low-damping layer to the high-damping layer, such that the precession in both layers can damp at the same rate [4]. As a result, the initial phase of the TR-MOKE signals also changes, which opens a negative or positive gap at high fields (> 10 kOe) for both modes, as shown in Fig. S2(d,f). This enables us to determine the difference between $\alpha_1$ and $\alpha_2$ by analyzing the initial phase of TR-MOKE signals.

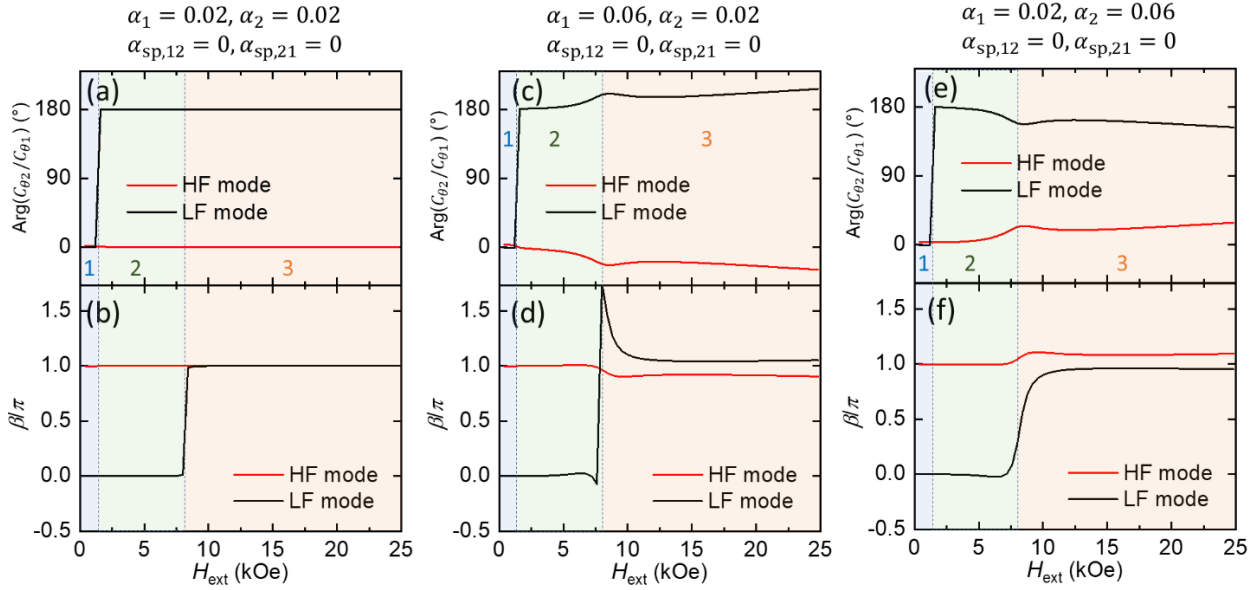

FIG. S2 Impact of $\alpha_1$ and $\alpha_2$ on the phase without mutual spin pumping. (a,c,e) The phase difference between the polar angles in two layers for HF and LF modes. (b,d,f) The calculated initial phase of TR-MOKE signals for each mode with $\alpha_1 = \alpha_2 = 0.02$ (a,b), $\alpha_1 = 0.06$ and $\alpha_2 = 0.02$ (c,d), and $\alpha_1 = 0.02$ and $\alpha_2 = 0.06$ (e,f). The mutual spin pumping is set as $\alpha_{sp,12} = \alpha_{sp,21} = 0$ for all three cases. The rest of the parameters used in this calculation can be found in TABLE SI.

The impact of mutual spin pumping on the precession phase is illustrated in Fig. S3, where three different cases of either the one-way ($\alpha_{sp,12}$ or $\alpha_{sp,21}$) or two-way (both $\alpha_{sp,12}$ and $\alpha_{sp,21}$) spin pumping are considered. A reference case without the consideration of mutual spin pumping ($\alpha_1 = 0.02$, $\alpha_2 = 0.06$, and $\alpha_{sp,12} = \alpha_{sp,21} = 0$) is also plotted (dashed curves) for the ease of comparison. In general, it can be seen that mutual spin pumping could also change the phase difference in the precession polar angles of two layers, and thus the initial phase of TR-MOKE signals noticeably. This can be explained by the damping modification resulting from spin pumping. In regions 2 and 3, Eq. (2) can be approximately rearranged as:

$$\frac{d\mathbf{m}_i}{dt} \approx -\gamma_i \mathbf{m}_i \times \mathbf{H}_{\text{eff},i} + \alpha_i \mathbf{m}_i \times \frac{d\mathbf{m}_i}{dt} - \frac{C_j}{C_i}\alpha_{\text{sp},ij}\cos(\theta_{0,1} - \theta_{0,2})\,\mathbf{m}_i \times \frac{d\mathbf{m}_i}{dt}$$

$$\approx -\gamma_i \mathbf{m}_i \times \mathbf{H}_{\text{eff},i} + \left[\alpha_i - \frac{C_j}{C_i}\alpha_{\text{sp},ij}\cos(\theta_{0,1} - \theta_{0,2})\right]\mathbf{m}_i \times \frac{d\mathbf{m}_i}{dt} \qquad (\text{S10})$$

$$= -\gamma_i \mathbf{m}_i \times \mathbf{H}_{\text{eff},i} + \bar{\alpha}_i \mathbf{m}_i \times \frac{d\mathbf{m}_i}{dt}$$

where $C_j/C_i$ represents the ratio of the cone angles in the $j$-th FM layer to the $i$-th FM layer. $C_j/C_i$ is positive for the in-phase mode and negative for the out-of-phase mode. $\theta_{0,1}$ and $\theta_{0,2}$ are the equilibrium polar angles of $\mathbf{M}_1$ and $\mathbf{M}_2$, as defined in Fig. 2(c). Therefore, the mutual spin-pumping term either enhances or reduces the damping depending on the mode. $\bar{\alpha}_i = \alpha_i - \frac{C_j}{C_i}\alpha_{\text{sp},ij}\cos(\theta_{0,1} - \theta_{0,2})$ represents the effective Gilbert damping in the $i$-th FM layer after considering the mutual spin-pumping effect. This modification to damping is more significant when the $i$-th layer is subservient with a smaller cone angle (*e.g.*, FM$_2$ for the HF mode in region 3), while the $j$-th layer is dominant with a much larger precession cone angle (*e.g.*, FM$_1$ for the LF mode in region 3), leading to a large ratio of $|C_j/C_i|$.

In Fig. S3(a), only the spin current injected from FM$_1$ to FM$_2$ is considered. According to the above analysis, $\alpha_{\text{sp},21}$ can only bring noticeable modifications to the damping of FM$_2$ when FM$_1$ is the dominant layer. Based on Fig. 3 in the main article, the LF mode in region 2 and HF mode in region 3 satisfy this condition (FM$_1$ dominant and FM$_2$ subservient). As shown in Fig. S3(a), the phase difference noticeably deviates from the reference case without mutual spin pumping (dashed curves) in region 2 for the LF mode (black curves) and in region 3 for the HF mode (red curves). For the LF mode in region 2, the precession motions in two layers are nearly out-of-phase (negative $C_1/C_2$); therefore, the spin pumping from FM$_1$ enhances the damping in FM$_2$. Since $\alpha_1$ (0.02) is less than $\alpha_2$ (0.06), the spin pumping from FM$_1$ to FM$_2$ further increases $|\bar{\alpha}_1 - \bar{\alpha}_2|$ between the two layers. Consequently, the phase difference shifts further away from 180°. While for the HF mode in region 3, $\alpha_{\mathrm{sp,21}}$ reduces the damping of FM$_2$ because $C_1/C_2$ is positive resulting from the near in-phase feature of this mode. Hence, $|\bar{\alpha}_1 - \bar{\alpha}_2|$ becomes smaller and the phase difference gets closer to 0°. In Fig. S3(c), only $\alpha_{\mathrm{sp,12}}$ is considered, which requires FM$_2$ as the dominant layer (the HF mode in region 2 and LF mode in region 3) for noticeable changes in $|\bar{\alpha}_1 - \bar{\alpha}_2|$. For the HF mode in region 2, spin pumping from FM$_2$ reduces $\bar{\alpha}_1$ given that the precession motions in two layers are nearly in phase (positive $C_2/C_1$). Therefore, $|\bar{\alpha}_1 - \bar{\alpha}_2|$ increases and the phase difference in Fig. S3(c) shifts further away from 0° in region 2. However, for the LF mode in regions 3, the nearly out-of-phase precession in two FM layers (negative $C_1/C_2$) increases $\bar{\alpha}_1$ and reduces $|\bar{\alpha}_1 - \bar{\alpha}_2|$. As a result, the phase difference in Fig. S3(c) shifts toward 180°. When both $\alpha_{\mathrm{sp,12}}$ and $\alpha_{\mathrm{sp,21}}$ are considered [Fig. S3(e)], a combined effect is expected for the phase difference with noticeable changes for both the HF and LF modes in regions 2 and 3.

The impacts of mutual spin pumping on the phase difference between the HF and LF modes are reflected by the initial phase of TR-MOKE signals [$\beta$ in Fig. S3(b,d,f)]. Compared with the reference case without mutual spin pumping (dashed curves), the introduction of mutual spin pumping tends to change the gap in $\beta$ between the two modes. As shown in Fig. S3(e,f), the values of two mutual-spin-pumping induced damping terms are chosen as $\alpha_{\mathrm{sp,12}} = 0.013$ and $\alpha_{\mathrm{sp,21}} = 0.004$, such that the $\beta$ gap of the initial phase of TR-MOKE signals is closed at high fields (region 3). Therefore, the initial phase of TR-MOKE signals provides certain measurement sensitivities to $\alpha_{\mathrm{sp,12}}$ and $\alpha_{\mathrm{sp,21}}$, which enables us to extract the values of $\alpha_{\mathrm{sp},ij}$ from measurement fitting. Here, we acknowledge that the measurement sensitivity to $\alpha_{\mathrm{sp},ij}$ from TR-MOKE is limited, which subsequently leads to relatively large error bars for $\alpha_{\mathrm{sp},ij}$ (see Table SI).

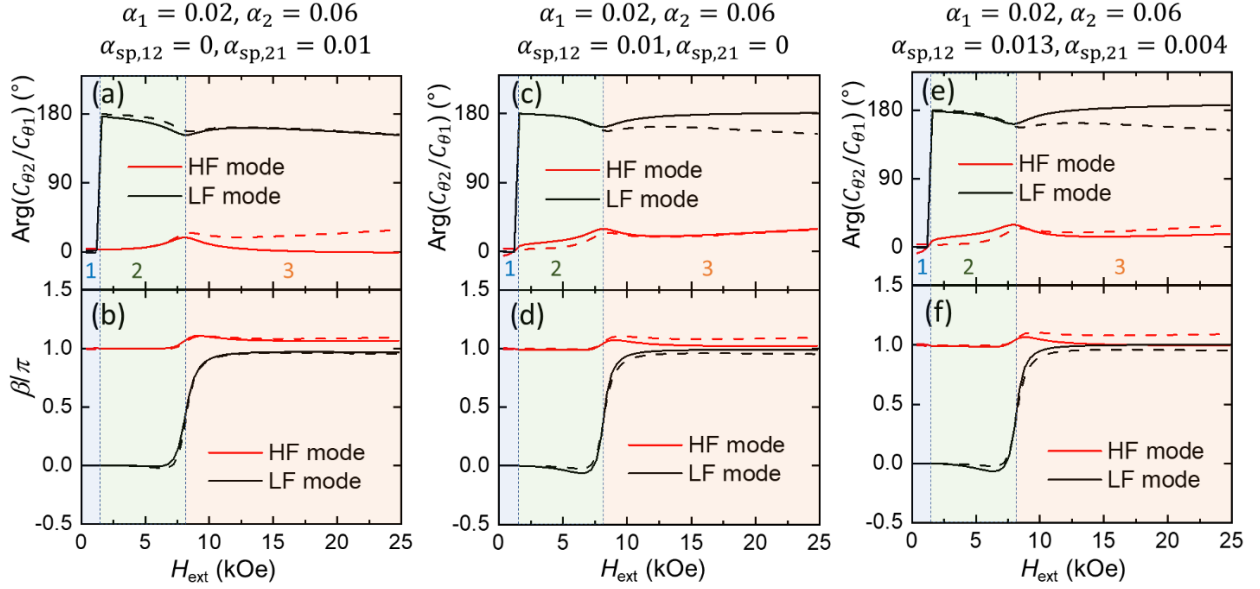

FIG. S3 Impact of mutual spin pumping on the phase with fixed damping values of $\alpha_1 = 0.02$ and $\alpha_2 = 0.06$. (a,c,e) The phase difference between the polar angles in two layers for HF and LF modes. (b,d,f) The calculated initial phase of TR-MOKE signals ($\beta$) for each mode with $\alpha_{sp,12} = 0$ and $\alpha_{sp,21} = 0.01$ (a,b), $\alpha_{sp,12} = 0.01$ and $\alpha_{sp,21} = 0$ (c,d), and $\alpha_{sp,12} = 0.013$ and $\alpha_{sp,21} = 0.004$ (e,f). For the third case (e,f), the values of mutual spin pumping are chosen to close the $\beta$ gap in panel (f) for $H_{ext} > 15$ kOe. The rest of the parameters used in this calculation can be found in TABLE SI. Dashed lines represent the reference case without mutual spin pumping ($\alpha_1 = 0.02$, $\alpha_2 = 0.06$, and $\alpha_{sp,12} = \alpha_{sp,21} = 0$).

**Supplemental Note 5: Region diagrams for p-SAFs with different degrees of asymmetries**

Figure S4 shows the region diagrams for p-SAFs with different degrees of asymmetries, represented by the difference of $H_{k,eff}$ in two FM layers. $H_{k,eff,1} = H_{k,eff,2}$ corresponds to the symmetric case (lowest asymmetry), as shown by Fig. S4(c). While the SAF in Fig. S4(a) has the highest asymmetry: $H_{k,eff,1} = 2$ kOe, $H_{k,eff,2} = 6$ kOe. Figure S4 clearly shows that $|\theta_{0,1} - \theta_{0,2}| > 90°$ is a necessary but not sufficient condition for region 1 (CW precession). Because regions 2 or 3 also appear to the left of the red curve (where $|\theta_{0,1} - \theta_{0,2}| > 90°$), especially when $\theta_H$ is close to 90° and $H_{k,eff,1}$ is close to $H_{k,eff,2}$.

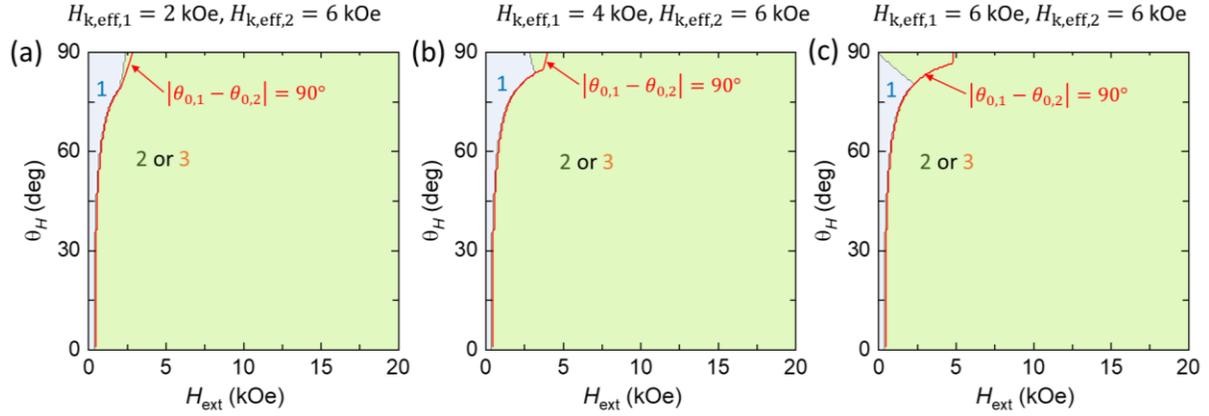

FIG. S4 Region diagrams of p-SAFs with different degrees of asymmetries: $H_{k,eff,1}$ = 2 kOe, $H_{k,eff,2}$ = 6 kOe (a), $H_{k,eff,1}$ = 4 kOe, $H_{k,eff,2}$ = 6 kOe (b), $H_{k,eff,1}$ = 6 kOe, $H_{k,eff,2}$ = 6 kOe (c). The blue background represents region 1. The green background covers regions 2 and 3. The red curve shows the conditions where $|\theta_{0,1} - \theta_{0,2}|$ = 90°. $|\theta_{0,1} - \theta_{0,2}|$ > 90° to the left of the red curve. $\alpha_1$, $\alpha_2$, $\alpha_{sp,12}$, and $\alpha_{sp,21}$ are set as zero. $\gamma_1 = \gamma_2$ = 17.8 rad ns$^{-1}$ kOe$^{-1}$. Values of the rest parameters are the same as those in Table SI.